\documentclass[12pt]{article}
\usepackage{graphicx}
\usepackage{amsmath}
\usepackage{amsbsy}
\usepackage{amssymb}
\usepackage{color}

\begin{document}

\title{\textsc{Two--mode optical tomograms: a possible experimental check of
the Robertson uncertainty relations}}
\author{V.I. Man'ko$^a$, G. Marmo$^b$, A. Simoni$^b$ and F. Ventriglia$^b$ \\
{\footnotesize \textit{$^a$P.N.Lebedev Physical Institute, Leninskii
Prospect 53, Moscow 119991, Russia}}\\
{\footnotesize {(e-mail: \texttt{manko@na.infn.it})}}\\
\textit{{\footnotesize {$^b$Dipartimento di Scienze Fisiche dell' Universit%
\`{a} ``Federico II" e Sezione INFN di Napoli,}}}\\
\textit{{\footnotesize {Complesso Universitario di Monte S. Angelo, via
Cintia, 80126 Naples, Italy}}}\\
{\footnotesize {(e-mail: \texttt{marmo@na.infn.it, simoni@na.infn.it,
ventriglia@na.infn.it})}}}
\maketitle

\begin{abstract}
The experimental check of two--mode Robertson uncertainty relations and
inequalities for highest quadrature moments is suggested by using homodyne
photon detection. The relation between optical tomograms and symplectic
tomograms is used to connect the tomographic dispersion matrix and the
quadrature components dispersion matrix of the two--mode field states.
\end{abstract}

\bigskip


\bigskip


\section{Introduction}

Recently \cite{UncertaintyASL,sol}, the possibility of experimentally checking of
the Schr\"{o}dinger--Robertson uncertainty relations \cite{Schr30,Rob29} for
conjugate position and momentum  by using homodyne photon detection 
was suggested.

The  homodyne photon detection yields the optical tomogram \cite%
{BerBer,VogelRis} of the photon quantum state and was implemented in the
available experiments \cite{Raymer,Mlynek,Lvovsky,Bellini,Solimeno} to reconstruct
the Wigner function of the state, which is interpreted as measuring the
quantum state.

In probability representation of quantum mechanics \cite%
{Mancini,Ventriglia1,Ventriglia2,Ventriglia3,manman} the state is identified
with the tomographic probability which is either an optical \cite%
{BerBer,VogelRis} or symplectic tomogram \cite{Mancini,D'Ariano97} or other
kinds of tomograms (see the review \cite{pedatom}).

In view of this, measuring the tomographic probability means to
measure the quantum state, since the tomogram contains all the information
on the quantum state including the information on the variances and
covariances of the conjugate variables (position and momentum or field
quadratures).

In this context, reconstructing the Wigner function for calculating other
state properties like the quadrature dispersion matrix is useless, since
all the characteristics can be obtained directly in terms of the measured
tomographic probability distributions.

The aim of this work is to suggest possible experiments to obtain optical
tomograms like for instance \cite{Solimeno}, where states of two--mode field
were studied by measuring one--mode field states tomogram, and to check
uncertainty relations for two--mode light.

The two--mode (multi--mode) uncertainty relations (see \cite%
{Rob29,Dodonov183,Kur,Sudars}) seem to have been never checked directly in
experiments.

Since the uncertainty relations are basic for quantum mechanics it is
reasonable to have both experimental confirmation of their validity and
knowledge of the experimental accuracy with which the uncertainty relations
are checked.

Up to now all the tomographic approaches to measure quantum states were
applied to one-mode light. Only recent experiments \cite{Solimeno} were
produced by homodyne detecting two--mode light. The experiments can be used
to detect presence or absence of entanglement phenomena in the field under
study.

In the two--mode case, the photon states which have been discussed the most have been assumed Gaussian. For Gaussian states the photon distribution function
was obtained in explicit form \cite{OlgaPhysRev94} both for the one-- and
multi--mode cases in terms of Hermite polynomials depending on many variables.

One of our aims is to discuss the photon distribution in a two--mode field by
means of the explicit expression given by the multi--variable Hermite
polynomials in \cite{OlgaPhysRev94} for Gaussian light. This makes possible
an additional control of both the Gaussianity property of the photon states
and the accuracy of the measurements. We devote particular attention to the uncertainty relations in the tomographic probability
representation for Gaussian states.

We also derive the tomographic form of the uncertainty
relations for arbitrary quantum states to get formulae where the Robertson
uncertainty relations are expressed in terms of two--mode symplectic
tomograms which generalize the results of \cite{UncertaintyASL} obtained for
one--mode states.

The paper is organized as follows. In section 2 we review the
Robertson uncertainty relations. In section 3, we briefly review the tomographic
probability representation of two--mode quantum states. In section 4, the
marginal tomograms of two--mode probability distributions are discussed and
the uncertainty relations for photon quadratures are given in the
tomographic form appropriate for experimental study. Inequalities for the
quadrature highest moments are given in section 5. Then state reconstruction
is discussed in section 6. The relations between quadratures and photon
statistics are considered in section 7. Conclusions and perspectives are
drawn in section 8.

\section{Two--mode uncertainty relations}

The Robertson \cite{Rob34} uncertainty relations for two--mode systems read
as a positivity condition for the matrix $\Sigma $ of the form:
\begin{equation}
\Sigma =%
\begin{pmatrix}
\sigma _{P_{1}P_{1}} & \sigma _{P_{1}P_{2}} & \sigma _{P_{1}Q_{1}} & \sigma
_{P_{1}Q_{2}} \\
\sigma _{P_{2}P_{1}} & \sigma _{P_{2}P_{2}} & \sigma _{P_{2}Q_{1}} & \sigma
_{P_{2}Q_{2}} \\
\sigma _{Q_{1}P_{1}} & \sigma _{Q_{1}P_{2}} & \sigma _{Q_{1}Q_{1}} & \sigma
_{Q_{1}Q_{2}} \\
\sigma _{Q_{2}P_{1}} & \sigma _{Q_{2}P_{2}} & \sigma _{Q_{2}Q_{1}} & \sigma
_{Q_{2}Q_{2}}%
\end{pmatrix}%
+\frac{\mathrm{i}}{2}%
\begin{pmatrix}
0 & 0 & -1 & 0 \\
0 & 0 & 0 & -1 \\
1 & 0 & 0 & 0 \\
0 & 1 & 0 & 0%
\end{pmatrix}%
.  \label{sigma}
\end{equation}%
The positivity condition for the above matrix
\begin{equation}
\Sigma \geq 0
\end{equation}%
means that all the principal minors of $\Sigma $ are non-negative. The
dispersion matrix contribution into the matrix $\Sigma $ is positive
(non-negative, in fact) both in classical and in quantum domain. The second
contribution is due to non-commutativity of the conjugate variables in the
quantum domain
\begin{equation}
Q_{j}P_{k}-P_{k}Q_{j}=i\delta _{jk}.
\end{equation}

Thus, one has the obvious inequalities
\begin{equation}
\sigma _{P_{k}P_{k}}\geq 0;\ \sigma _{Q_{k}Q_{k}}\geq 0;\ k=1,2,
\end{equation}
accompanied by the Schr\"{o}dinger--Robertson \cite{Rob29,Schr30}
inequalities for each mode:
\begin{equation}
\sigma _{P_{k}P_{k}}\sigma _{Q_{k}Q_{k}}-\sigma _{P_{k}Q_{k}}^{2}-\frac{1}{4}%
\geq 0;\ k=1,2.
\end{equation}

Besides, there are inequalities which are cubic in variances and
covariances, as:
\begin{equation}
\det
\begin{pmatrix}
\sigma _{P_{1}P_{1}} & \sigma _{P_{1}P_{2}} & \sigma _{P_{1}Q_{1}}-\frac{\mathrm{i}}{2%
} \\
\sigma _{P_{2}P_{1}} & \sigma _{P_{2}P_{2}} & \sigma _{P_{2}Q_{1}} \\
\sigma _{Q_{1}P_{1}}+\frac{\mathrm{i}}{2} & \sigma _{Q_{1}P_{2}} & \sigma
_{Q_{1}Q_{1}}%
\end{pmatrix}%
\geq 0.
\end{equation}

Also, one has a quartic inequality which is equivalent to the non-negativity
of the $4-$th principal minor of $\Sigma :$%
\begin{equation}
\det
\begin{pmatrix}
\sigma _{P_{1}P_{1}} & \sigma _{P_{1}P_{2}} & \sigma _{P_{1}Q_{1}} & \sigma
_{P_{1}Q_{2}} \\
\sigma _{P_{2}P_{1}} & \sigma _{P_{2}P_{2}} & \sigma _{P_{2}Q_{1}} & \sigma
_{P_{2}Q_{2}} \\
\sigma _{Q_{1}P_{1}} & \sigma _{Q_{1}P_{2}} & \sigma _{Q_{1}Q_{1}} & \sigma
_{Q_{1}Q_{2}} \\
\sigma _{Q_{2}P_{1}} & \sigma _{Q_{2}P_{2}} & \sigma _{Q_{2}Q_{1}} & \sigma
_{Q_{2}Q_{2}}%
\end{pmatrix}
\geq \frac{1}{16}.
\end{equation}

These inequalities can be checked by measuring the matrix elements of the
dispersion matrix. This can be done in the tomographic approach.

\section{Tomograms -- symplectic and optical}

The two--mode symplectic tomogram was introduced in \cite{D'Ariano97}.
Let us consider two homodyne quadratures
\begin{equation}
\hat{X}_{1}=\mu _{1}Q_{1}+\nu _{1}P_{1};\ \hat{X}_{2}=\mu _{2}Q_{2}+\nu
_{2}P_{2}\ .  \label{homoquad}
\end{equation}
where the $Q$'s and the $P$'s are the usual position and momentum operators.
The symplectic tomogram of a two mode density state $\hat{\rho}\left(
1,2\right) $ depends on six real variables and reads: \cite{pedatom}
\begin{equation}
\mathcal{M}\left( X_{1},\mu _{1},\nu _{1};X_{2},\mu _{2},\nu _{2}\right) =%
\mathrm{Tr}\left[ \hat{\rho}\left( 1,2\right) \delta \left( X_{1}\hat{I}-%
\hat{X}_{1}\right) \delta \left( X_{2}\hat{I}-\hat{X}_{2}\right) \right] \ .
\end{equation}
In terms of Wigner function $W$ the symplectic tomogram reads:
\begin{eqnarray}
&&\mathcal{M}\left( X_{1},\mu _{1},\nu _{1};X_{2},\mu _{2},\nu _{2}\right) =
\\
&&\int W\left( q_{1},p_{1};q_{2},p_{2}\right) \delta \left( X_{1}-\mu
_{1}q_{1}-\nu _{1}p_{1}\right) \delta \left( X_{2}-\mu _{2}q_{2}-\nu
_{2}p_{2}\right) \frac{dq_{1}dp_{1}dq_{2}dp_{2}}{4\pi ^{2}}\ .  \notag
\end{eqnarray}
and it is nonnegative and normalized:
\begin{equation}
\int \mathcal{M}\left( X_{1},\mu _{1},\nu _{1};X_{2},\mu _{2},\nu
_{2}\right) dX_{1}dX_{2}=1\ .
\end{equation}

For
\begin{equation}
\mu _{1}=\cos \theta _{1},\ \nu _{1}=\sin \theta _{1};\ \mu _{2}=\cos \theta
_{2},\ \nu _{2}=\sin \theta _{2}
\end{equation}%
one gets the optical two--mode tomogram depending on four essential real
variables:
\begin{eqnarray}
&&\mathcal{W}\left( X_{1},\theta _{1};X_{2},\theta _{2}\right) = \\
&&\mathcal{M}\left( X_{1},\mu _{1}=\cos \theta _{1},\nu _{1}=\sin \theta
_{1};X_{2},\mu _{2}=\cos \theta _{2},\nu _{2}=\sin \theta _{2}\right) .
\notag
\end{eqnarray}%
This tomogram is the joint probability distribution of $X_{1}$\ and $X_{2}.$%
\ In view of this one has
\begin{eqnarray}
\mathcal{W}^{(1)}\left( X_{1},\theta _{1}\right) &=&\int \mathcal{W}\left(
X_{1},\theta _{1};X_{2},\theta _{2}\right) dX_{2}; \\
\mathcal{W}^{(2)}\left( X_{2},\theta _{2}\right) &=&\int \mathcal{W}\left(
X_{1},\theta _{1};X_{2},\theta _{2}\right) dX_{1}\ ,  \notag
\end{eqnarray}%
where $\mathcal{W}^{(1)}\left( X_{1},\theta _{1}\right) $\ and $\mathcal{W}%
^{(2)}\left( X_{2},\theta _{2}\right) $ are the optical tomograms of the
first and second mode, respectively. In fact, $\mathcal{W}\left(
X_{1},\theta _{1};X_{2},\theta _{2}\right)$ is a function of four variables,
and integration over one variable would be expected to yield a function of
three variables. On the contrary, the above formulae show that integration
of the two--mode tomogram over a random variable $X_k$ gives a function
independent of the associated variable parameter $\theta_k.$ However, this
property seems obvious in view of the physical meaning of a tomogram as
joint probability density of two random position variables, measured in new
reference frames in phase space, rotated by angles $\theta_1,\theta_2$.
Since the tomogram of the first mode state is a marginal of the joint
probability distribution, the integration over the second mode position
washes out any information about the reference frame where the integrated
position was measured. Such property takes place also for other tomographic
probability distributions like spin tomograms of multi--qudit states \cite%
{pedatom}.

It is worthy to note that for the Wigner function of a two--mode field,
which is only a quasi--distribution function, such a property does not hold,
so that one has to integrate over both conjugate position and momentum to
obtain a one--mode Wigner function:
\begin{eqnarray}
W^{(1)}\left( q_{1},p_{1}\right) &=&\int W\left(
q_{1},p_{1};q_{2},p_{2}\right) \frac{dq_{2}dp_{2}}{2\pi }, \\
W^{(2)}\left( q_{2},p_{2}\right) &=&\int W\left(
q_{1},p_{1};q_{2},p_{2}\right) \frac{dq_{1}dp_{1}}{2\pi }.  \notag
\end{eqnarray}

The one--mode tomograms $\mathcal{W}^{(k)}$'s are related to the
corresponding Wigner functions of the states as
\begin{equation*}
\mathcal{W}^{(k)}\left( X_{k},\theta _{k}\right) =\int W^{(k)}\left(
q_{k},p_{k}\right) \delta \left( X_{k}-\cos \theta _{k}q_{k}-\sin \theta
_{k}p_{k}\right) \frac{dq_{k}dp_{k}}{2\pi }\ ,
\end{equation*}%
with $k=1,2.$

From Eq.$\left( \ref{homoquad}\right) $ one obtains:
\begin{equation}
\hat{X}_{k}^{2}=\mu _{k}^{2}Q_{k}^{2}+\nu _{k}^{2}P_{k}^{2}+2\mu _{k}\nu
_{k}\left( \frac{Q_{k}P_{k}+P_{k}Q_{k}}{2}\right) ;\ \left( k=1,2\right) \ ,
\label{rel1}
\end{equation}%
and
\begin{equation}
\hat{X}_{k}\hat{X}_{j}=\mu _{k}\mu _{j}Q_{k}Q_{j}+\nu _{k}\nu
_{j}P_{k}P_{j}+\mu _{k}\nu _{j}Q_{k}P_{j}+\nu _{k}\mu _{j}P_{k}Q_{j}\ ;\
\left( j\neq k=1,2\right) \ .  \label{rel2}
\end{equation}

Due to the physical meaning of the tomogram as probability distribution, for
homodyne quadratures one has
\begin{equation}
\mathrm{Tr}\left[ \hat{\rho}\left( 1,2\right) \hat{X}_{k}^{n}\right] =\int
X_{k}^{n}\mathcal{W}^{(k)}\left( X_{k},\theta _{k}\right) dX_{k}\ ;\ \left(
n=0,1,2,3,\ldots \right) \ ,  \label{moment}
\end{equation}
and
\begin{equation}
\mathrm{Tr}\left[ \hat{\rho}\left( 1,2\right) \hat{X}_{k}\hat{X}_{j}\right]
=\int X_{k}X_{j}\mathcal{W}\left( X_{1},\theta _{1};X_{2},\theta _{2}\right)
dX_{1}dX_{2}\ ;\ \left( j,k=1,2\right) \ .  \label{cov}
\end{equation}

Bearing in mind the relations $\left( \ref{rel1}\right) ,\left( \ref{rel2}%
\right) $, by means of the optical tomograms $\mathcal{W}\left( X_{1},\theta
_{1};X_{2},\theta _{2}\right) ,\mathcal{W}^{(1)}\left( X_{1},\theta
_{1}\right) ,\mathcal{W}^{(2)}\left( X_{2},\theta _{2}\right) $ one can
express the matrix elements of the dispersion matrix of $\Sigma $ (eq. $%
\left( \ref{sigma}\right) $) in terms of integrals on the\ right hand side
of the above equations $\left( \ref{moment}\right) ,\left( \ref{cov}\right) $%
. In fact one gets:
\begin{equation}
\sigma _{Q_{k}Q_{k}}=\int X_{k}^{2}\mathcal{W}^{(k)}\left( X_{k},\theta
_{k}=0\right) dX_{k}-\left( \int X_{k}\mathcal{W}^{(k)}\left( X_{k},\theta
_{k}=0\right) dX_{k}\right) ^{2}\ ,
\end{equation}
and
\begin{equation}
\sigma _{P_{k}P_{k}}=\int X_{k}^{2}\mathcal{W}^{(k)}\left( X_{k},\theta _{k}=%
\frac{\pi }{2}\right) dX_{k}-\left( \int X_{k}\mathcal{W}^{(k)}\left(
X_{k},\theta _{k}=\frac{\pi }{2}\right) dX_{k}\right) ^{2}\ ,
\end{equation}
with $k=1,2.$ Moreover:
\begin{equation}
\sigma _{Q_{k}P_{k}}=\sigma _{X_{k}X_{k}}\left( \frac{\pi }{4}\right) -\frac{%
1}{2}\sigma _{X_{k}X_{k}}\left( 0\right) -\frac{1}{2}\sigma
_{X_{k}X_{k}}\left( \frac{\pi }{2}\right) \ .
\end{equation}

One can see that it is possible to measure in terms of tomograms all the
highest moments of the quadratures
\begin{equation}
\left\langle Q_{k}^{n}P_{k}^{m}\right\rangle =\mathrm{Tr}\left[ \hat{\rho}%
\left( 1,2\right) Q_{k}^{n}P_{k}^{m}\right] ;\ \left( k=1,2\right) \ .
\end{equation}%
For example, in the case of cubic moments, one has
\begin{equation}
\left\langle Q_{k}^{3}\right\rangle =\int X_{k}^{3}\mathcal{W}^{(k)}\left(
X_{k},\theta _{k}=0\right) dX_{k};\ \left\langle P_{k}^{3}\right\rangle
=\int X_{k}^{3}\mathcal{W}^{(k)}\left( X_{k},\theta _{k}=\frac{\pi }{2}%
\right) dX_{k}\ .
\end{equation}%
By using
\begin{eqnarray}
\hat{X}_{k}^{3} &=&\mu _{k}^{3}Q_{k}^{3}+\nu _{k}^{3}P_{k}^{3}+\mu
_{k}^{2}\nu _{k}\left( Q_{k}^{2}P_{k}+Q_{k}P_{k}Q_{k}+P_{k}Q_{k}^{2}\right)
\\
&+&\mu _{k}\nu _{k}^{2}\left(
P_{k}^{2}Q_{k}+P_{k}Q_{k}P_{k}+Q_{k}P_{k}^{2}\right)  \notag
\end{eqnarray}%
and the commutation relations of the quadratures, one obtains
\begin{equation}
Q_{k}P_{k}Q_{k}=P_{k}Q_{k}^{2}+iQ_{k}\ ;\
Q_{k}^{2}P_{k}=P_{k}Q_{k}^{2}+2iQ_{k}\
\end{equation}%
and similarly
\begin{equation}
P_{k}Q_{k}P_{k}=P_{k}^{2}Q_{k}+iP_{k}\ ;\
Q_{k}P_{k}^{2}=P_{k}^{2}Q_{k}+2iP_{k}
\end{equation}%
so that
\begin{equation}
\hat{X}_{k}^{3}=\mu _{k}^{3}Q_{k}^{3}+\nu _{k}^{3}P_{k}^{3}+3\mu _{k}^{2}\nu
_{k}\left( P_{k}Q_{k}^{2}+iQ_{k}\right) +3\mu _{k}\nu _{k}^{2}\left(
P_{k}^{2}Q_{k}+iP_{k}\right) \ .
\end{equation}

\section{Marginals of two--mode tomograms}

Let us consider light modes which can be obtained by means of optical
devices from two initial ones described by operators $a$ and $b$ satisfying
the commutation relations
\begin{equation}
\left[ a,a^{\dagger }\right] =\left[ b,b^{\dagger }\right] =1\ ;\ \left[ a,b%
\right] =\left[ a,b^{\dagger }\right] =\left[ a^{\dagger },b\right] =0.
\end{equation}
Then one can make symplectic transformations and get the modes described by
the operators
\begin{eqnarray}
c &=&\frac{1}{\sqrt{2}}\left( a+b\right) ;\ d=\frac{1}{\sqrt{2}}\left(
a-b\right) ;  \label{FourModes} \\
e &=&\frac{1}{\sqrt{2}}\left( a+\mathrm{i}b\right) ;\ f=\frac{1}{\sqrt{2}}\left(
a-\mathrm{i}b\right) .  \notag
\end{eqnarray}
One can readily check that
\begin{equation}
\left[ c,c^{\dagger }\right] =\left[ d,d^{\dagger }\right] =\left[
e,e^{\dagger }\right] =\left[ f,f^{\dagger }\right] =1.
\end{equation}

The initial modes can be expressed in terms of quadratures:
\begin{equation}
a=\frac{1}{\sqrt{2}}\left( Q_{1}+\mathrm{i}P_{1}\right) ;b=\frac{1}{\sqrt{2}}\left(
Q_{2}+\mathrm{i}P_{2}\right)
\end{equation}
Besides, one has homodyne quadrature operators for each of these modes
\begin{equation}
\hat{X}_{a}\left( \mu _{1},\nu _{1}\right) =\left( \mu _{1}Q_{1}+\nu
_{1}P_{1}\right) ;\ \hat{X}_{b}\left( \mu _{2,}\nu _{2}\right) =\mu
_{2}Q_{2}+\nu _{2}P_{2}\ .  \label{homquad}
\end{equation}
where one can take local oscillator phases to set
\begin{equation}
\mu _{1}=\cos \theta _{1},\ \nu _{1}=\sin \theta _{1};\ \mu _{2}=\cos \theta
_{2},\ \nu _{2}=\sin \theta _{2}\ .
\end{equation}
Then one can consider the homodyne quadrature operators \ for all four modes
given by $\left( \ref{FourModes}\right) $ in terms of initial quadratures
operators
\begin{eqnarray}
\hat{X}_{c}\left( \mu _{3},\nu _{3}\right) &=&\frac{1}{2}\mu _{3}\left(
Q_{1}+Q_{2}\right) +\frac{1}{2}\nu _{3}\left( P_{1}+P_{2}\right)
\label{FourQuad} \\
\hat{X}_{d}\left( \mu _{4},\nu _{4}\right) &=&\frac{1}{2}\mu _{4}\left(
Q_{1}-Q_{2}\right) +\frac{1}{2}\nu _{4}\left( P_{1}-P_{2}\right)  \notag \\
\hat{X}_{e}\left( \mu _{5},\nu _{5}\right) &=&\frac{1}{2}\mu _{5}\left(
Q_{1}-P_{2}\right) +\frac{1}{2}\nu _{5}\left( Q_{2}+P_{1}\right)  \notag \\
\hat{X}_{f}\left( \mu _{6},\nu _{6}\right) &=&\frac{1}{2}\mu _{6}\left(
Q_{1}+P_{2}\right) +\frac{1}{2}\nu _{6}\left( P_{1}-Q_{2}\right)  \notag
\end{eqnarray}
We now shift from labels $a,b,c,d,e,f$ to labels $1,2,3,4,5,6,$
respectively, so that $\hat{X}_{a}\left( \mu _{1},\nu _{1}\right)
\rightarrow \hat{X}_{1}\left( \mu _{1},\nu _{1}\right) ,$ $\hat{X}_{b}\left(
\mu _{2,}\nu _{2}\right) \rightarrow \hat{X}_{2}\left( \mu _{2,}\nu
_{2}\right) $ and so on. In the above equations the parameters may be chosen
as
\begin{equation*}
\mu _{k}=\cos \theta _{k},\ \nu _{k}=\sin \theta _{k},\ \left(
k=3,4,5,6\right) .
\end{equation*}
The above equations $\left( \ref{FourQuad}\right) $\ can be given a vector
form:
\begin{equation}
\boldsymbol{\hat{X}}=S\boldsymbol{Q\ },
\end{equation}
where the vector $\boldsymbol{\hat{X}},\boldsymbol{Q\ }$ have operator
components $\hat{X}_{3},\hat{X}_{4},\hat{X}_{5},\hat{X}_{6}$ and $%
P_{1},P_{2},Q_{1},Q_{2}$ respectively, while the matrix $S$ reads
\begin{equation}
S=\frac{1}{2}
\begin{pmatrix}
\nu _{3} & \nu _{3} & \mu _{3} & \mu _{3} \\
\nu _{4} & -\nu _{4} & \mu _{4} & -\mu _{4} \\
\nu _{5} & -\mu _{5} & \mu _{5} & \nu _{5} \\
\nu _{6} & \mu _{6} & \nu _{6} & -\mu _{6}%
\end{pmatrix}%
\end{equation}
and is invertible, as $\det S\neq 0$ in the generic case. So one can
solve with respect to $\boldsymbol{Q}$ and obtain
\begin{equation}
\boldsymbol{Q}=S^{-1}\boldsymbol{\hat{X}}
\end{equation}
or, taking mean values:
\begin{equation}
\left\langle \boldsymbol{Q}\right\rangle =S^{-1}\left\langle \boldsymbol{%
\hat{X}}\right\rangle
\end{equation}
where operators are averaged in the density state $\rho :\left\langle \hat{A}%
\right\rangle :=\mathrm{Tr}\left( \rho \hat{A}\right) $ .

One-mode homodyne detectors can measure the six optical tomograms
\begin{equation}
\mathcal{W}_{k}\left( X_{k},\theta _{k}\right) =\mathrm{Tr}\left( \rho \hat{X%
}_{k}\right) ,\left( k=1,2,\ldots ,6\right) .
\end{equation}
For each $k-$mode one has the Schr\"{o}dinger--Robertson inequality
expressed in terms of measured tomograms \cite{UncertaintyASL} associated
with the two--mode light state
\begin{eqnarray}
F(\theta _{k}) &=&\left( \int X_{k}^{2}\mathcal{W}_{k}(X_{k},\theta
_{k})dX_{k}-\left[ \int X_{k}\mathcal{W}_{k}(X_{k},\theta _{k})dX_{k}\right]
^{2}\right) \times  \label{F(Theta)} \\
&&\left( \int X_{k}^{2}\mathcal{W}_{k}(X_{k},\theta _{k}+\frac{\pi }{2}%
)dX_{k}-\left[ \int X_{k}\mathcal{W}_{k}(X_{k},\theta _{k}+\frac{\pi }{2}%
)dX_{k}\right] ^{2}\right)  \notag \\
&&-\left\{ \int X_{k}^{2}\mathcal{W}_{k}(X_{k},\theta _{k}+\frac{\pi }{4}%
)dX_{k}-\left[ \int X_{k}\mathcal{W}_{k}(X_{k},\theta _{k}+\frac{\pi }{4}%
)dX_{k}\right] ^{2}\right.  \notag \\
&&-\frac{1}{2}\left[ \int X_{k}^{2}\mathcal{W}_{k}(X_{k},\theta _{k})dX_{k}-%
\left[ \int X_{k}\mathcal{W}_{k}(X_{k},\theta _{k})dX_{k}\right] ^{2}\right.
\notag \\
&&+\left. \left. \int X_{k}^{2}\mathcal{W}_{k}(X_{k},\theta _{k}+\frac{\pi }{%
2})dX_{k}-\left[ \int X_{k}\mathcal{W}_{k}(X_{k},\theta _{k}+\frac{\pi }{2}%
)dX_{k}\right] ^{2}\right] \right\}^2 -\frac{1}{4}\geq 0.  \notag
\end{eqnarray}
From the homodyne quadrature operators $\left( \ref{homquad}\right) $ one
gets:
\begin{equation}
Q_{1}=\hat{X}_{1}\left( 1,0\right) ,\ P_{1}=\hat{X}_{1}\left( 0,1\right)
;Q_{2}=\hat{X}_{2}\left( 1,0\right) ,\ P_{2}=\hat{X}_{2}\left( 0,1\right) ;
\end{equation}
Moreover:
\begin{eqnarray}
Q_{1}^{2} &=&\hat{X}_{1}^{2}\left( 1,0\right) ,\ P_{1}^{2}=\hat{X}%
_{1}^{2}\left( 0,1\right) ; \\
\frac{1}{2}\left\{ Q_{1},P_{1}\right\} &=&\hat{X}_{1}^{2}\left( \frac{\sqrt{2%
}}{2},\frac{\sqrt{2}}{2}\right) -\frac{1}{2}\left[ \hat{X}_{1}^{2}\left(
1,0\right) +\hat{X}_{1}^{2}\left( 0,1\right) \right]  \notag
\end{eqnarray}
and analogously, turning label $1$ into label $2:$%
\begin{eqnarray}
Q_{2}^{2} &=&\hat{X}_{2}^{2}\left( 1,0\right) ,\ P_{2}^{2}=\hat{X}%
_{2}^{2}\left( 0,1\right) ; \\
\frac{1}{2}\left\{ Q_{2},P_{2}\right\} &=&\hat{X}_{2}^{2}\left( \frac{\sqrt{2%
}}{2},\frac{\sqrt{2}}{2}\right) -\frac{1}{2}\left[ \hat{X}_{2}^{2}\left(
1,0\right) +\hat{X}_{2}^{2}\left( 0,1\right) \right] .  \notag
\end{eqnarray}
Also:
\begin{eqnarray}
Q_{1}Q_{2} &=&2\hat{X}_{3}^{2}\left( 1,0\right) -\frac{1}{2}\left[ \hat{X}%
_{1}^{2}\left( 1,0\right) +\hat{X}_{2}^{2}\left( 1,0\right) \right] ;
\label{quaddiag} \\
P_{1}P_{2} &=&2\hat{X}_{3}^{2}\left( 0,1\right) -\frac{1}{2}\left[ \hat{X}%
_{1}^{2}\left( 0,1\right) +\hat{X}_{2}^{2}\left( 0,1\right) \right] ,  \notag
\end{eqnarray}
and finally
\begin{eqnarray}
Q_{1}P_{2} &=&-2\hat{X}_{5}^{2}\left( 1,0\right) +\frac{1}{2}\left[ \hat{X}%
_{1}^{2}\left( 1,0\right) +\hat{X}_{2}^{2}\left( 0,1\right) \right] ;
\label{quadcross} \\
Q_{2}P_{1} &=&2\hat{X}_{5}^{2}\left( 0,1\right) -\frac{1}{2}\left[ \hat{X}%
_{1}^{2}\left( 0,1\right) +\hat{X}_{2}^{2}\left( 1,0\right) \right] .  \notag
\end{eqnarray}

The obtained equalities allow to express the Robertson uncertainty relations
in terms of homodyne tomograms which are experimentally measured. There are
also relations which are compatible with the properties of the different
homodyne quadratures $\left( \ref{homquad}\right) ,\left( \ref{FourQuad}%
\right) ,$ for example
\begin{equation}
\hat{X}_{3}\left( 1,0\right) =\frac{1}{2}\left[ \hat{X}_{1}\left( 1,0\right)
+\hat{X}_{2}\left( 1,0\right) \right]
\end{equation}
and many other including quadrature equalities.

The expressions for variances and covariances of the two--mode field
quadrature components are, with $k=1,2:$%
\begin{eqnarray}
\sigma _{Q_{k}Q_{k}} &=&\left( \int X_{k}^{2}\mathcal{W}_{k}(X_{k},\theta
_{k}=0)dX_{k}-\left[ \int X_{k}\mathcal{W}_{k}(X_{k},\theta _{k}=0)dX_{k}%
\right] ^{2}\right) ;  \notag \\
\sigma _{P_{k}P_{k}} &=&\left( \int X_{k}^{2}\mathcal{W}_{k}(X_{k},\theta
_{k}=\frac{\pi }{2})dX_{k}-\left[ \int X_{k}\mathcal{W}_{k}(X_{k},\theta
_{k}=\frac{\pi }{2})dX_{k}\right] ^{2}\right) ;  \notag \\
\sigma _{Q_{k}P_{k}} &=&\left( \int X_{k}^{2}\mathcal{W}_{k}(X_{k},\theta
_{k}=\frac{\pi }{4})dX_{k}-\left[ \int X_{k}\mathcal{W}_{k}(X_{k},\theta
_{k}=\frac{\pi }{4})dX_{k}\right] ^{2}\right)  \notag \\
&-&\frac{1}{2}\left( \sigma _{Q_{k}Q_{k}}+\sigma _{P_{k}P_{k}}\right) .
\end{eqnarray}
The same expressions appear in formula $\left( \ref{F(Theta)}\right) $\
defining $F(\theta _{k}).$ Besides, in view of $\left( \ref{quaddiag}\right)
$ one has
\begin{eqnarray}
\sigma _{Q_{1}Q_{2}} &=&2\left( \int X_{3}^{2}\mathcal{W}_{3}(X_{3},\theta
_{3}=0)dX_{3}-\left[ \int X_{3}\mathcal{W}_{3}(X_{3},\theta _{3}=0)dX_{3}%
\right] ^{2}\right)  \notag \\
&-&\frac{1}{2}\left( \sigma _{Q_{1}Q_{1}}+\sigma _{Q_{2}Q_{2}}\right) ;
\notag \\
\sigma _{P_{1}P_{2}} &=&2\left( \int X_{3}^{2}\mathcal{W}_{3}(X_{3},\theta
_{3}=\frac{\pi }{2})dX_{3}-\left[ \int X_{3}\mathcal{W}_{3}(X_{3},\theta
_{3}=\frac{\pi }{2})dX_{3}\right] ^{2}\right)  \notag \\
&-&\frac{1}{2}\left( \sigma _{P_{1}P_{1}}+\sigma _{P_{2}P_{2}}\right) ,
\end{eqnarray}
and finally, from $\left( \ref{quadcross}\right) $ one gets
\begin{eqnarray}
\sigma _{Q_{1}P_{2}} &=&-2\left( \int X_{5}^{2}\mathcal{W}_{5}(X_{5},\theta
_{5}=0)dX_{5}-\left[ \int X_{5}\mathcal{W}_{5}(X_{5},\theta _{5}=0)dX_{5}%
\right] ^{2}\right)  \notag \\
&+&\frac{1}{2}\left( \sigma _{Q_{1}Q_{1}}+\sigma _{P_{2}P_{2}}\right) ;
\notag \\
\sigma _{Q_{2}P_{1}} &=&2\left( \int X_{5}^{2}\mathcal{W}_{5}(X_{5},\theta
_{5}=\frac{\pi }{2})dX_{5}-\left[ \int X_{5}\mathcal{W}_{5}(X_{5},\theta
_{5}=\frac{\pi }{2})dX_{5}\right] ^{2}\right)  \notag \\
&-&\frac{1}{2}\left( \sigma _{Q_{2}Q_{2}}+\sigma _{P_{1}P_{1}}\right) .
\end{eqnarray}

Inserting the obtained variances and covariances into the matrix $\Sigma $
defined in Eq.$\left( \ref{sigma}\right) $ one can calculate all its
principal minors in terms of the measured tomograms $\mathcal{W}%
_{k}(X_{k},\theta _{k})$ and check by direct experimental data the Robertson
uncertainty relations by means of the positivity of such minors. One can
tell that the experimental data for all six modes are redundant to get the
dispersion matrix for the quadratures. One could use other mode
combinations. For example, we did not use the mode data associated with the
tomogram $\mathcal{W}_{6}(X_{6},\theta _{6}).$ Another set of tomograms
could be used to get the dispersion matrix. This variety is useful to make
extra control of accuracy of the measurements because the dispersion must be
the same irrespectively of the particular set of tomograms used. Thus one
has the dispersion matrix with \textquotedblleft
commutator\textquotedblright contributions in different permutations of
basis, e.g.
\begin{equation*}
\Sigma =%
\begin{pmatrix}
\sigma _{P_{1}P_{1}} & \sigma _{P_{1}P_{2}} & \sigma _{P_{1}Q_{1}}-\frac{\mathrm{i}}{2%
} & \sigma _{P_{1}Q_{2}} \\
\sigma _{P_{2}P_{1}} & \sigma _{P_{2}P_{2}} & \sigma _{P_{2}Q_{1}} & \sigma
_{P_{2}Q_{2}}-\frac{\mathrm{i}}{2} \\
\sigma _{Q_{1}P_{1}}+\frac{\mathrm{i}}{2} & \sigma _{Q_{1}P_{2}} & \sigma
_{Q_{1}Q_{1}} & \sigma _{Q_{1}Q_{2}} \\
\sigma _{Q_{2}P_{1}} & \sigma _{Q_{2}P_{2}}+\frac{\mathrm{i}}{2} & \sigma
_{Q_{2}Q_{1}} & \sigma _{Q_{2}Q_{2}}%
\end{pmatrix}%
\end{equation*}%
and
\begin{equation*}
\Sigma ^{\prime }=%
\begin{pmatrix}
\sigma _{Q_{1}Q_{1}} & \sigma _{Q_{1}Q_{2}} & \sigma _{Q_{1}P_{1}}+\frac{\mathrm{i}}{2%
} & \sigma _{Q_{1}P_{2}} \\
\sigma _{Q_{2}Q_{1}} & \sigma _{Q_{2}Q_{2}} & \sigma _{Q_{2}P_{1}} & \sigma
_{Q_{2}P_{2}}+\frac{\mathrm{i}}{2} \\
\sigma _{P_{1}Q_{1}}-\frac{\mathrm{i}}{2} & \sigma _{P_{1}Q_{2}} & \sigma
_{P_{1}P_{1}} & \sigma _{P_{1}P_{2}} \\
\sigma _{P_{2}Q_{1}} & \sigma _{P_{2}Q_{2}}-\frac{\mathrm{i}}{2} & \sigma
_{P_{2}P_{1}} & \sigma _{P_{2}P_{2}}%
\end{pmatrix}%
.
\end{equation*}%
All the elements of the matrices are expressed in terms of measured
tomograms. One has to check the non--negativity of the principal minors.
Although theoretically the non--negativity of the principal minors found in
one basis induces their nonnegativity in all of the other bases, the
checking of the experimental data looks different since the order of the
checking depends on the basis. For example, in the case of the matrix $%
\Sigma $ the second principal minor provides a checking of the Schr\"{o}%
dinger-Robertson uncertainty relations, while using $\Sigma ^{\prime }$ the
second leading principal minor
\begin{equation}
M_{2}=\sigma _{Q_{1}Q_{1}}\sigma _{Q_{2}Q_{2}}-\sigma _{Q_{1}Q_{2}}^{2}\geq 0
\end{equation}%
one checks a classical property which does not contain Planck's constant
influence. Thus, to check the uncertainty relations and to control the
accuracy of the data, it seems to be reasonable to check the non--negativity
of the principal minors in all the bases.

The non--negativity of the third leading principal minor yields the
constraints obtained from the matrices $\Sigma $ and $\Sigma ^{\prime }$
respectively:
\begin{equation*}
\det
\begin{pmatrix}
\sigma _{P_{1}P_{1}} & \sigma _{P_{1}P_{2}} & \sigma _{P_{1}Q_{1}}-\frac{\mathrm{i}}{2%
} \\
\sigma _{P_{2}P_{1}} & \sigma _{P_{2}P_{2}} & \sigma _{P_{2}Q_{1}} \\
\sigma _{Q_{1}P_{1}}+\frac{\mathrm{i}}{2} & \sigma _{Q_{1}P_{2}} & \sigma
_{Q_{1}Q_{1}}%
\end{pmatrix}%
\geq 0
\end{equation*}%
and
\begin{equation*}
\det
\begin{pmatrix}
\sigma _{Q_{1}Q_{1}} & \sigma _{Q_{1}Q_{2}} & \sigma _{Q_{1}P_{1}}+\frac{\mathrm{i}}{2%
} \\
\sigma _{Q_{2}Q_{1}} & \sigma _{Q_{2}Q_{2}} & \sigma _{Q_{2}P_{1}} \\
\sigma _{P_{1}Q_{1}}-\frac{\mathrm{i}}{2} & \sigma _{P_{1}Q_{2}} & \sigma
_{P_{1}P_{1}}%
\end{pmatrix}%
\geq 0.
\end{equation*}%
The last minor is just the determinant of the matrix $\Sigma ,$ or $\Sigma
^{\prime }.$

\section{Measuring highest moments of quadratures by homodyne detector}

Let us discuss first how to measure highest moments for one-mode light, say
the $a-$mode given by Eq.$(\ref{homquad}),$ $\hat{X}_{1}\left( \mu _{1},\nu
_{1}\right) =\mu _{1}Q_{1}+\nu _{1}P_{1}$. Mean values, variances and
covariances are given in terms of the tomogram $\mathcal{W}_{1}\left(
X_{1},\theta _{1}\right) .$ Let us construct cubic moments. Then, dropping
label $1,$ one has to find the moments of the operators $%
P^{3},P^{2}Q,PQ^{2},Q^{3}$ because the remaining may be expressed by
commutators as
\begin{eqnarray}
PQP &=&PPQ+P\left[ Q,P\right] =P^{2}Q+\mathrm{i}P\Rightarrow \left\langle
PQP\right\rangle =\left\langle P^{2}Q\right\rangle +\mathrm{i}\left\langle
P\right\rangle , \\
QPP &=&PQP+\left[ Q,P\right] P=P^{2}Q+2\mathrm{i}P\Rightarrow \left\langle
QP^{2}\right\rangle =\left\langle P^{2}Q\right\rangle +2\mathrm{i}\left\langle
P\right\rangle ,  \notag
\end{eqnarray}
and analogously
\begin{eqnarray}
\left\langle QPQ\right\rangle &=&\left\langle PQ^{2}\right\rangle
+\mathrm{i}\left\langle Q\right\rangle , \\
\left\langle Q^{2}P\right\rangle &=&\left\langle PQ^{2}\right\rangle
+2\mathrm{i}\left\langle Q\right\rangle .  \notag
\end{eqnarray}

The cubic power $\hat{X}^{3}\left( \mu ,\nu \right) $ reads
\begin{eqnarray}
\hat{X}^{3}\left( \mu ,\nu \right) &=& \mu ^{3}Q^{3}+\nu ^{3}P^{3}+\mu
^{2}\nu \left( Q^{2}P+QPQ+PQ^{2}\right)  \notag \\
&+&\mu \nu ^{2}\left( P^{2}Q+PQP+QP^{2}\right)
\end{eqnarray}
so that
\begin{eqnarray}
\left\langle \hat{X}^{3}\right\rangle \left( \mu ,\nu \right) &=&\mu
^{3}\left\langle Q^{3}\right\rangle +\nu ^{3}\left\langle P\right\rangle
^{3}+3\mu ^{2}\nu \left( \left\langle PQ^{2}\right\rangle +i\left\langle
Q\right\rangle \right)  \notag \\
&+&3\mu \nu ^{2}\left( \left\langle P^{2}Q\right\rangle +i\left\langle
P\right\rangle \right) .
\end{eqnarray}
The means of the quadratures read
\begin{equation}
\left\langle \hat{X}\right\rangle \left( 1,0\right) =\left\langle
Q\right\rangle \ ,\ \left\langle \hat{X}\right\rangle \left( 0,1\right)
=\left\langle P\right\rangle .
\end{equation}
Besides, one has
\begin{equation}
\left\langle \hat{X}^{3}\right\rangle \left( 1,0\right) =\left\langle
Q^{3}\right\rangle \ ,\ \left\langle \hat{X}^{3}\right\rangle \left(
0,1\right) =\left\langle P^{3}\right\rangle ,
\end{equation}
and
\begin{eqnarray}
\left\langle \hat{X}^{3}\right\rangle \left( \mu ,\nu \right) &=&\mu
^{3}\left\langle \hat{X}^{3}\right\rangle \left( 1,0\right) +\nu
^{3}\left\langle \hat{X}^{3}\right\rangle \left( 0,1\right) +3\mu ^{2}\nu
\left( \left\langle PQ^{2}\right\rangle +i\left\langle \hat{X}\right\rangle
\left( 1,0\right) \right)  \notag \\
&+&3\mu \nu ^{2}\left( \left\langle P^{2}Q\right\rangle +i\left\langle \hat{X%
}\right\rangle \left( 0,1\right) \right) .
\end{eqnarray}
Introducing the function
\begin{eqnarray}
A\left( \mu ,\nu \right) &:&=\left\langle \hat{X}^{3}\right\rangle \left(
\mu ,\nu \right) -\mu ^{3}\left\langle \hat{X}^{3}\right\rangle \left(
1,0\right) -\nu ^{3}\left\langle \hat{X}^{3}\right\rangle \left( 0,1\right)
\notag \\
&&-3\mu ^{2}\nu i\left\langle \hat{X}\right\rangle \left( 1,0\right) -3\mu
\nu ^{2}i\left\langle \hat{X}\right\rangle \left( 0,1\right)
\end{eqnarray}
we obtain two linear equations for the remaining two moments:
\begin{eqnarray}
A\left( \mu _{\alpha },\nu _{\alpha }\right) &=&3\mu _{\alpha }^{2}\nu
_{\alpha }\left\langle PQ^{2}\right\rangle +3\mu _{\alpha }\nu _{\alpha
}^{2}\left\langle P^{2}Q\right\rangle ; \\
A\left( \mu _{\beta },\nu _{\beta }\right) &=&3\mu _{\beta }^{2}\nu _{\beta
}\left\langle PQ^{2}\right\rangle +3\mu _{\beta }\nu _{\beta
}^{2}\left\langle P^{2}Q\right\rangle ;  \notag
\end{eqnarray}
which can be readily solved in terms of the homodyne quadratures given by
the tomogram $\mathcal{W}\left( X,\theta \right) $ only.

The previous construction of the solutions
\begin{equation}
\left\langle PQ^{2}\right\rangle =\frac{1}{\Delta }\det
\begin{pmatrix}
A\left( \mu _{\alpha },\nu _{\alpha }\right) & 3\mu _{\alpha }\nu _{\alpha
}^{2} \\
A\left( \mu _{\beta },\nu _{\beta }\right) & 3\mu _{\beta }\nu _{\beta }^{2}%
\end{pmatrix}
;\left\langle P^{2}Q\right\rangle =\frac{1}{\Delta }\det
\begin{pmatrix}
3\mu _{\alpha }^{2}\nu _{\alpha } & A\left( \mu _{\alpha },\nu _{\alpha
}\right) \\
3\mu _{\beta }^{2}\nu _{\beta } & A\left( \mu _{\beta },\nu _{\beta }\right)%
\end{pmatrix}
,  \label{alfabeta}
\end{equation}
with
\begin{equation}
\Delta =\det
\begin{pmatrix}
3\mu _{\alpha }^{2}\nu _{\alpha } & 3\mu _{\alpha }\nu _{\alpha }^{2} \\
3\mu _{\beta }^{2}\nu _{\beta } & 3\mu _{\beta }\nu _{\beta }^{2}%
\end{pmatrix}
,
\end{equation}
shows that the same procedure can be applied to get all the highest moments $%
\left\langle P^{n}Q^{m}\right\rangle $ and $\left\langle
P^{m}Q^{n}\right\rangle \ \left( n,m=0,1,\ldots \right) $ in terms of the
tomogram $\mathcal{W}\left( X,\theta \right) $ only$.$ It provides the tool
to check all the known high moments quantum uncertainty relations \cite%
{Dodonov183} in fact both in one mode and multi--mode case. As an example we
derive simple uncertainty relations for cubic moments.

Let us consider the linear forms:
\begin{equation}
\hat{f}=y_{1}Q+y_{2}P^{2}\ ;\ \hat{f}^{\dagger }=y_{1}^{\ast }Q+y_{2}^{\ast
}P^{2}\ .
\end{equation}%
The obvious inequality for the mean value
\begin{equation}
\left\langle \hat{f}\hat{f}^{\dagger }\right\rangle \geq 0
\end{equation}%
gives a condition of nonnegativity for the quadratic form
\begin{equation}
y_{1}y_{1}^{\ast }\left\langle Q^{2}\right\rangle +y_{1}y_{2}^{\ast
}\left\langle QP^{2}\right\rangle +y_{2}y_{1}^{\ast }\left\langle
P^{2}Q\right\rangle +y_{2}y_{2}^{\ast }\left\langle P^{4}\right\rangle \geq
0.
\end{equation}%
Thus the matrix of the quadratic form
\begin{equation}
M=%
\begin{pmatrix}
\left\langle Q^{2}\right\rangle  & \left\langle QP^{2}\right\rangle  \\
\left\langle P^{2}Q\right\rangle  & \left\langle P^{4}\right\rangle
\end{pmatrix}%
\end{equation}%
must be nonnegative, and this implies
\begin{equation}
\left\langle Q^{2}\right\rangle \left\langle P^{4}\right\rangle
-\left\langle QP^{2}\right\rangle \left\langle P^{2}Q\right\rangle \geq 0.
\label{ThirdOrdIneq}
\end{equation}%
This inequality can be written in terms of tomograms as
\begin{eqnarray}
\int X^{2}\mathcal{W}(X,\theta =0)dX\ \int X^{4}\mathcal{W}(X,\theta =\frac{%
\pi }{2})dX-\left[ \left\langle QP^{2}\right\rangle \left\langle
P^{2}Q\right\rangle \right] _{\theta _{\alpha },\theta _{\beta }}\geq 0
\end{eqnarray}%
where local oscillator phases, for instance $\theta _{\alpha }=\pi /3,\theta
_{\beta }=2\pi /3,$ are taken \`{\i}n Eq. $\left( \ref{alfabeta}\right) $,
so that the parameters $\left( \mu _{\alpha },\nu _{\alpha }\right) $ and $%
\left( \mu _{\beta },\nu _{\beta }\right) $ are $\left( \sqrt{3}%
/2,1/2\right) $ and $\left( 1/2,\sqrt{3}/2\right) $ respectively. Of course,
one could use other suitable local oscillator phases, such that $\Delta \neq
0$ in Eq. $\left( \ref{alfabeta}\right) .$ The above cubic-in-quadrature
uncertainty relation must be satisfied by any of the six modes used in
experiments \cite{Solimeno}.

In view of the generalization proposed in \cite{UncertaintyASL} for Schr\"{o}dinger--Robertson uncertainty relations,
an analogous generalization can be proposed for the above highest order moments inequality, that can be written in covariant form, i.e. for all the local oscillator phases as:
\begin{eqnarray}
&&\int X^{2}\mathcal{W}(X,\theta)dX\ \int X^{4}\mathcal{W}(X,\theta +\frac{%
\pi }{2})dX \notag \\
&& -\left[ \left\langle QP^{2}\right\rangle \left\langle
P^{2}Q\right\rangle \right] _{\theta + \theta _{\alpha },\theta + \theta _{\beta }}\geq 0,
\label{covariantineq}
\end{eqnarray}
where, as before, Eq. $\left( \ref{alfabeta}\right) $ has to be used with the new values of local oscillator phases, say ${\theta + \pi /3,\theta + 2 \pi /3}.$

\section{State reconstruction}

The one--mode measurement can be used to get complete information on the
two--mode state. In fact, the complete information is contained in the
symplectic tomogram $\mathcal{M}\left( X_{1},\mu _{1},\nu _{1};X_{2},\mu
_{2},\nu _{2}\right) $ or in the optical tomogram $\mathcal{W}\left(
X_{1},\theta _{1};X_{2},\theta _{2}\right) $. The same information on the
state is contained in the Wigner function $W\left(
q_{1},p_{1};q_{2},p_{2}\right) .$

We recall that one--mode states have tomogram
\begin{equation}
\mathcal{W}^{(1)}\left( X_{1},\theta _{1}\right) =\int \mathcal{W}\left(
X_{1},\theta _{1};X_{2},\theta _{2}\right) dX_{2}
\end{equation}
and Wigner function
\begin{equation}
W^{(1)}\left( q_{1},p_{1}\right) =\int W\left(
q_{1},p_{1};q_{2},p_{2}\right) \frac{dq_{2}dp_{2}}{2\pi }.
\end{equation}

A natural question to ask: is it possible to find the two--mode state either
in terms of tomogram or in terms of Wigner function, if only one--mode
tomograms can be measured? The answer is positive, in fact. It was shown
\cite{VentriWigAut} that a symplectic transformation $V$ acts on Wigner function
or tomogram of the anti-transformed state $V\rho V^{\dagger }.$ In
experiment \cite{Solimeno} the one-mode tomograms are measured for such
symplectically transformed states. Thus one has the marginal probability
distributions depending on a set of parameters sufficient to find out the
two--mode state tomogram (Wigner function or density operator).

This property may be understood in general terms. The quadratures of multi--mode field
close on the Lie algebra of the Weyl--Heisenberg group. All the highest quadrature
moments are determined by the elements of the enveloping of this 
Weyl--Heisenberg algebra. Then, any new basis of the Lie algebra obtained by
the initial one by a linear invertible transformation gives rise to the
same enveloping algebra. So, having new transformed modes and measuring the
corresponding one--mode tomograms, makes possible to find out all the highest
moments of the multi--mode field. This means that one can reconstruct the
multi--mode state tomogram by measuring only a set of suitably chosen
one--mode tomograms. We now show this procedure on the example of a
two--mode field.

The idea of reconstructing the two--mode density operator by measuring
several one--mode density operators is the following one. The marginals
give us the possibility to find all the quadrature moments of two--mode
light in terms of one-mode tomograms only. Then the Wigner function,
tomogram or density operator are expressed in terms of moments. For example
the characteristic function of the tomogram, which is the Fourier transform
of the tomogram, reads
\begin{eqnarray}
\mathcal{\tilde{W}}\left( K_{1},\theta _{1};K_{2},\theta _{2}\right)
&&:=\int \mathcal{W}\left( X_{1},\theta _{1};X_{2},\theta _{2}\right) \exp %
\left[ \mathrm{i}\left( K_{1}X_{1}+K_{2}X_{2}\right) \right] dX_{1}dX_{2}  \notag \\
&&=\sum\limits_{n,m=0}^{\infty }\frac{\left( \mathrm{i}K_{1}\right) ^{n}\left(
\mathrm{i}K_{2}\right) ^{m}}{n!m!}\left\langle X_{1}^{n}X_{2}^{m}\right\rangle \left(
\theta _{1},\theta _{2}\right)
\end{eqnarray}%
where the moments explicitly are:
\begin{equation}
\left\langle X_{1}^{n}X_{2}^{m}\right\rangle \left( \theta _{1},\theta
_{2}\right) =\int X_{1}^{n}X_{2}^{m}\mathcal{W}\left( X_{1},\theta
_{1};X_{2},\theta _{2}\right) dX_{1}dX_{2}.
\end{equation}%
The knowledge of all moments allows the reconstruction of the characteristic
function. Then the tomogram is given by a Fourier anti--transform:.
\begin{equation}
\mathcal{W}\left( X_{1},\theta _{1};X_{2},\theta _{2}\right) =\int \mathcal{%
\tilde{W}}\left( K_{1},\theta _{1};K_{2},\theta _{2}\right) \exp \left[
-\mathrm{i}\left( K_{1}X_{1}+K_{2}X_{2}\right) \right] \frac{dK_{1}dK_{2}}{\left(
2\pi \right) ^{2}}.
\end{equation}

Analogously, the Wigner function of the two--mode state is known as soon as
the moments
\begin{equation}
\left\langle q_{1}^{n}p_{1}^{m}q_{2}^{n^{\prime }}p_{2}^{m^{\prime
}}\right\rangle =\int q_{1}^{n}p_{1}^{m}q_{2}^{n^{\prime }}p_{2}^{m^{\prime
}}W\left( q_{1},p_{1};q_{2},p_{2}\right) \frac{dq_{1}dp_{1}dq_{2}dp_{2}}{%
\left( 2\pi \right) ^{2}}
\end{equation}%
are known. Again, these moments determine the characteristic function:
\begin{eqnarray}
&&\tilde{W}\left( \xi _{1},\eta _{1};\xi _{2},\eta _{2}\right)  \notag \\
&:=&\int W\left( q_{1},p_{1};q_{2},p_{2}\right) \exp \left[ \mathrm{i}\left( \xi
_{1}q_{1}+\eta _{1}p_{1}+\xi _{2}q_{2}+\eta _{2}p_{2}\right) \right] \frac{%
dq_{1}dp_{1}dq_{2}dp_{2}}{\left( 2\pi \right) ^{2}}  \notag \\
&=&\sum\limits_{n,m,n^{\prime },m^{\prime }=0}^{\infty }\frac{\left( \mathrm{i}\xi
_{1}\right) ^{n}\left( \mathrm{i}\eta _{1}\right) ^{m}\left(\mathrm{i}\xi _{2}\right)
^{n^{\prime }}\left( \mathrm{i}\eta _{2}\right) ^{m^{\prime }}}{n!m!n^{\prime
}!m^{\prime }!}\left\langle q_{1}^{n}p_{1}^{m}q_{2}^{n^{\prime
}}p_{2}^{m^{\prime }}\right\rangle
\end{eqnarray}%
and the Wigner function is obtained by anti--Fourier transforming:
\begin{eqnarray}
&&W\left( q_{1},p_{1};q_{2},p_{2}\right) \\
&&=\int \tilde{W}\left( \xi _{1},\eta _{1};\xi _{2},\eta _{2}\right) \exp %
\left[ -\mathrm{i}\left( \xi _{1}q_{1}+\eta _{1}p_{1}+\xi _{2}q_{2}+\eta
_{2}p_{2}\right) \right] \frac{d\xi _{1}d\eta _{1}d\xi _{2}d\eta _{2}}{%
\left( 2\pi \right) ^{2}}.  \notag
\end{eqnarray}%
As one can see all the moments can be obtained by measuring the one--mode
optical tomograms as it was shown in examples of quadratic and cubic
moments. The procedure to find other highest moments is iterative and
starting from the mean values $\left\langle Q_{k}\right\rangle ,\left\langle
P_{k}\right\rangle ,\left( k=1,2\right) ,$ the variances and covariances one
obtains all the highest moments by the measured one-mode tomograms of the
two--mode $a$ and $b$ light and its symplectically transformed
modes $(a\pm b)/\sqrt{2},(a \pm\mathrm{i}b)/\sqrt{2}.$ Thus
the state of two--mode light is described by one tomogram depending on two
random quadratures $X_{1},X_{2}$ or by a sufficient set of one--mode
tomograms $\mathcal{W}_{k}\left( X_{k},\theta _{k}\right) $ which are
appropriate marginals.

\section{Photon statistics}

One can formulate the problem of measuring state in terms of photon
statistical properties of the measured two--mode light. For Gaussian states
the photon statistics is described by multi--variable Hermite polynomials.
For small number of photons, the expressions can be easily constructed. The
photon distribution are determined by the highest moments
\begin{equation}
\left\langle \hat{n}_{1}^{^{k_{1}}}\hat{n}_{2}^{^{k_{2}}}\right\rangle =%
\mathrm{Tr}\left( \rho \hat{n}_{1}^{^{k_{1}}}\hat{n}_{2}^{^{k_{2}}}\right) ,
\end{equation}%
where
\begin{equation}
\hat{n}_{1}=\frac{1}{2}\left( Q_{1}^{2}+P_{1}^{2}-1\right) \ ;\ \hat{n}_{2}=%
\frac{1}{2}\left( Q_{2}^{2}+P_{2}^{2}-1\right) .
\end{equation}%
Thus, measuring the photon statistics implies to measure the photon
quadrature highest moments. Since highest moments satisfy the quantum
uncertainty relations, the photon statistics (quantum correlations)
demonstrate difference with properties of classical electromagnetic field. The photon statistics can be studied using
measured optical (symplectic) tomograms. Experimental check of the quantum
uncertainty relations serves not only to investigate the degree of accuracy
with which nowadays the uncertainty relations are known to be fulfilled. Since there
are no doubts that the quantum mechanics is a correct theory and the
uncertainty relations must be fulfilled, the results of the experiments can
serve also to control the correctness of the experimental tools used in
homodyne detecting photon states. There exist inequalities in which the
highest moments of quadrature components are involved (see, e.g., the review
\cite{Dodonov183}). One can reformulate these highest order inequalities in
terms of tomographic quadrature moments given for example in Eq. $(\ref%
{ThirdOrdIneq})$ and to obtain extra inequalities expressed in terms of the
experimental values of the optical tomogram. Moreover, we suggested the possibility to use the covariant form of Eq. $(\ref%
{ThirdOrdIneq})$, given by Eq. (\ref{covariantineq}), more suitable for an experimental check. The tomographic probability
approach can be applied also for two--mode and multi--mode photon states
especially for Gaussian states for which their properties like photon
statistics are sufficiently known.

Thus, the photon distribution function for the two--mode field is explicitly
given in \cite{OlgaPhysRev94} in terms of Hermite polynomials of four
variables, related to quadrature variances and covariances of the Gaussian
field states.

Thus, measuring both photon statistics and optical tomograms provides the
possibility of a cross checking of the quantum inequalities for the quadrature
highest moments.

\section{Conclusions}
To summarize, we list the main results of this paper.
For two--mode quantum field we express the photon quadrature uncertainty
relations, like the Robertson's ones, in terms of measurable optical
tomograms of one--mode quantum electromagnetic field. We suggest to use the
given tomographic expression of the Robertson's inequality to control the
accuracy of the homodyne photon state detection. Also, we give examples of
inequalities for highest moments of the photon quadratures for one--mode field.
 We have expressed  all the
inequalities in tomographic form, in particular in covariant form, this is suitable for experimental
checking. Such checking, to the best of our knowledge, has not yet been done
due to the absence of the technique appropriate for the experimental
verification of these basic inequalities, which can be violated in the
classical domain. We have connected the checking of the photon statistics
to the possible suggested experimental checking of the quadrature
statistics. The generalization of the tomographic approach to study
Robertson uncertainty relations to the multi--mode field is shown to be
straightforward.

\end{document}